\documentstyle[eqsecnum,preprint,prd,aps]{revtex}
\newcommand{\spect}[3]{{}^{#1}\!#2_{#3}}

\begin{document}
\thispagestyle{empty}
\draft
\title{Relativistic two-photon and two-gluon decay rates \\of heavy quarkonia}
\author{Suraj N. Gupta and James M. Johnson}
\address{ Department of Physics, Wayne State University, Detroit,
	Michigan 48202}
\author{Wayne W. Repko}
\address{Department of Physics and Astronomy, Michigan State University,
	East Lansing, Michigan 48824}
\date{April 8, 1996}
\maketitle
\begin{abstract}

	The decay rates of $c\bar{c}$ and $b\bar{b}$ through 
two-photon or two-gluon annihilations are obtained by using totally
relativistic decay amplitudes and a sophisticated
quantum-chromodynamic potential model for heavy quarkonia.  Our results
for the photonic and gluonic widths of the $\spect{1}{S}{0}$,
$\spect{3}{P}{0}$, and the $\spect{3}{P}{2}$ states are in excellent
agreement with the available experimental data.  The procedures and
mathematical techniques used by us for the treatment of the
fermion-antifermion bound states are also applicable to other decay
processes.
\end{abstract}

\clearpage
\setlength{\parindent}{24pt}

\section{INTRODUCTION}

	The decay rates of heavy quarkonia through two-photon or 
two-gluon annihilations were first obtained in the nonrelativistic
approximation \cite{appel,barb}, and found to be inadequate in the
light of the experimental data \cite{olsson}.  Improvements to the
earlier results,
therefore, have been explored by various authors by including the
relativistic corrections \cite{suaya,berg,li,ackleh}.  The quarkonium decay
is usually treated by a suitable adaptation of the matrix element for
the annihilation of a free quark-antiquark pair, and this treatment is
either supported by appealing to the Bethe-Salpeter approach with
instantaneous approximation \cite{berg} or simply regarded as an
artifice \cite{ackleh}.

	In view of the ambiguity involved in the treatment of the 
quarkonium decays, we shall obtain the decay rates by following two
different approaches based on different assumptions regarding the role
of the potential energy in a bound state.  In both approaches, fully
relativistic matrix elements for the quark-antiquark annihilation will
be used.  Moreover, unlike the earlier authors, we shall use a
realistic quarkonium potential model \cite{quark} which has proved
highly successful in the investigation of the $c\bar{c}$ and $b\bar{b}$
spectra with spin splittings.  Our results for the decays of the
$\spect{1}{S}{0}$, $\spect{3}{P}{0}$, and $\spect{3}{P}{2}$ states of
$c\bar{c}$ and $b\bar{b}$ will be compared with the earlier results
of other authors as well as with the available
experimental data.

	The general procedure and its applications to the $S$ and $P$ 
states of the fermion-antifermion bound states are described in Sec.~II
and III, while the quarkonium photonic and gluonic widths according to
two different approaches are obtained in Secs.~IV and V, which is
followed by a discussion of our results in Sec.~VI.

\section{Fermion-Antifermion Bound-State Decays}

	Let us first consider the annihilation of a pair of an electron 
and a positron of four-momenta $p$ and $q$ into two photons of
four-momenta $k$ and $k'$.  The second-order contribution of the
scattering operator for this process is
\begin{equation}
S=V^{-2}(2\pi)^4 \delta(p+q-k-k') F \; a_{\bf e}^\ast({\bf k}) 
	a_{\bf e'}^\ast({\bf k'}) b_s({\bf q}) a_r({\bf p}) ,
\end{equation}
where
\begin{equation}
F=\frac{ie^2}{2} \frac{1}{(k_0 k'_0)^{1/2}}\; \bar{v}_s({\bf q})\left[
	{\bf e'}\cdot\bbox{\gamma}\frac{i(p-k)\cdot \gamma-m}{(p-k)^2+m^2}
		{\bf e}\cdot\bbox{\gamma}
	+{\bf e}\cdot\bbox{\gamma}\frac{i(p-k')\cdot \gamma-m}{(p-k')^2+m^2}
		{\bf e'}\cdot\bbox{\gamma}\right] u_r({\bf p}).
\end{equation}
It is to be noted that
\begin{eqnarray}
\bar{v}_s({\bf q})(iq\cdot\gamma-m)=0,& & \qquad (ip\cdot\gamma+m)
	u_r({\bf p})=0,\nonumber \\
p^2=q^2=-m^2, & &\qquad k^2=k'^2=0,
\end{eqnarray}
so that $F$ can be simplified as
\begin{equation}
F=\frac{ie^2}{2} \frac{1}{(k_0 k'_0)^{1/2}} \;\bar{v}_s({\bf q})\left[
	{\bf e}'\cdot\bbox{\gamma} \frac{2i{\bf p}\cdot{\bf e}-i(k\cdot 
		\gamma)({\bf e}\cdot\bbox{\gamma})}{-2{\bf k}\cdot{\bf p}}
	+{\bf e}\cdot\bbox{\gamma} \frac{2i{\bf p}\cdot{\bf e'}-i(k'\cdot 
		\gamma)({\bf e'}\cdot\bbox{\gamma})}{-2{\bf k'}\cdot{\bf p}}
	\right] u_r({\bf p}).
\end{equation}

	It is possible to convert the matrix element from the 
Dirac form to Pauli form without making any approximation by the
substitutions\cite{sng}
\begin{equation}
u_r({\bf p})=\left(\frac{m+p_0}{2p_0}\right)^{1/2}
	\left(\begin{array}{c}1\\ \displaystyle\frac{\bbox{\sigma}
	\cdot{\bf p}}{m+p_0}\end{array}\right)\alpha_r,\qquad
v_s({\bf q})=\left(\frac{m+q_0}{2q_0}\right)^{1/2}
	\left(\begin{array}{c}\displaystyle\frac{\bbox{\sigma}
	\cdot{\bf q}}{m+q_0}\\ 1\end{array}\right)\beta_s,
\end{equation}
and
\begin{equation}
\gamma_i=\left(\begin{array}{cc}0&-i\sigma_i\\i\sigma_i&0
	\end{array}\right),\qquad
\gamma_4=\left(\begin{array}{cc}1&0\\0&-1\end{array}\right),
\end{equation}
together with the charge-conjugation relation
\begin{equation}
\beta_s^\ast=\alpha_s^T i\sigma_2.
\end{equation}

	Then, after reducing the products of the $\sigma$~matrices, it
is found that
\begin{equation}
F=\alpha^T_s i\sigma_2 O \alpha_r  \label{amp1}
\end{equation}
with
\begin{eqnarray}
O=\frac{ie^2}{2k_0^2[(\hat{\bf k}\cdot{\bf p})^2-p_0^2]}
& & \Biggl[ imk_0(\hat{\bf k}\cdot{\bf e}\times{\bf e'})  
	 -(\bbox{\sigma}\cdot\hat{\bf k})({\bf e}\cdot
		{\bf e'})(\hat{\bf k}\cdot{\bf p})k_0\nonumber\\
& &\quad -[(\bbox{\sigma}\cdot{\bf e})({\bf p}\cdot{\bf e'})
	+(\bbox{\sigma}\cdot{\bf e'})({\bf p}\cdot{\bf e})]p_0
		\nonumber\\
& &\quad -[(\bbox{\sigma}\cdot{\bf e})({\bf p}\cdot{\bf e'})
	-(\bbox{\sigma}\cdot{\bf e'})({\bf p}\cdot{\bf e})]
		\frac{(\hat{\bf k}\cdot{\bf p})(k_0-p_0)}{p_0}\nonumber\\
& &\quad \left.+(\bbox{\sigma}\cdot{\bf p})\frac{k_0({\bf e}
	\cdot{\bf e'})(\hat{\bf k}\cdot{\bf p})^2+2p_0({\bf e}
		\cdot{\bf p})({\bf e'}\cdot{\bf p})}{p_0(m+p_0)}\right],
\end{eqnarray}
where we have used the center-of-mass relations
\begin{equation}
{\bf q}=-{\bf p},\quad q_0=p_0,\quad {\bf k'}=-{\bf k},\quad k'_0=k_0,
\end{equation}
as well as
\begin{equation}
{\bf k}=k_0\hat{\bf k},\quad \hat{\bf k}\cdot{\bf e}=
	\hat{\bf k}\cdot{\bf e'}=0,
\end{equation}
but avoided the use of the energy conservation relation
\begin{equation}
p_0=k_0. \label{altenergy1}
\end{equation}

	Now, let $\Psi$ denote a positronium wave function, 
which can be Fourier decomposed as
\begin{equation}
\Psi({\bf r})=\frac{1}{(2\pi)^{3/2}}\int d{\bf p} \Psi({\bf p})
	e^{i{\bf p}\cdot{\bf x}}.
\end{equation}
Following the usual approach, we assume that the decay amplitude 
for the ${\bf p}$ component of the positronium wave function can be
obtained from the free electron-positron annihilation amplitude
(\ref{amp1}) by ignoring the relation (\ref{altenergy1}), and treating
$p_0$ as a variable given by
\begin{equation}
p_0=(m^2+{\bf p}^2)^{1/2}, \quad 0< |{\bf p}|<\infty,\label{energy}
\end{equation}
while
\begin{equation}
k_0=\frac{1}{2}m_{e\bar{e}}.
\end{equation}
This assumption implies that the effect of the bound-state potential
energy is simply to nullify the energy conservation relation for the
free-state annihilation amplitude.

	Consequently, the amplitude for positronium decay into two
photons is
\begin{equation}
\bar{F}=\frac{1}{(2\pi)^{3/2}} \int d{\bf p} F \Psi({\bf p}), 
	\label{amp2}
\end{equation}
and the resulting decay rate is given by
\begin{equation}
\Gamma(e\bar{e}\rightarrow\gamma\gamma)=\frac{1}{(2\pi)^2}
	\int d\Omega_{\bf k}\frac{k_0^2}{2}|\bar{F}|^2 . \label{gamma}
\end{equation}

	Furthermore, the decay rates for quarkonia
can be obtained by using the quarkonium wave functions, setting
$k_0=\frac{1}{2}M_{Q\bar{Q}}$, and making the usual multiplicative
replacements in (\ref{gamma}).  For the decay rate
$\Gamma(Q\bar{Q}\rightarrow\gamma\gamma)$, the replacement is
\begin{equation}
\alpha^2\rightarrow Ne_Q^4 \alpha^2, \quad N=3,\label{photonic}
\end{equation}
while $\Gamma(Q\bar{Q}\rightarrow gg)$ can be obtained from
$\Gamma(Q\bar{Q}\rightarrow\gamma\gamma)$ by the replacement
\begin{equation}
e_Q^4\alpha^2\rightarrow \frac{2}{9}\alpha_s^2 .\label{gluonic}
\end{equation}

\section{$S$ and $P$ State Decay Rates}

	We shall apply the treatment of Sec.~II to obtain the 
decay rates for those $S$ and $P$ states of positronium which can decay
into two photons, and for this purpose we shall use the wave functions
in the matrix representation, given in Appendix~A.

\subsection{$\spect{1}{S}{0}$ decay}

	For the $\spect{1}{S}{0}$ state (\ref{1s0}), the decay 
amplitude, given by (\ref{amp2}) and (\ref{amp1}), takes the form
\begin{equation}
\bar{F}=\frac{i}{(2\pi)^{3/2}}\int d{\bf p} \frac{1}{(8\pi)^{1/2}} 
	Tr[O] \phi(p).
\end{equation}
Since terms linear in $\bbox{\sigma}$ in $O$ do not contribute 
to the trace, $\bar{F}$ reduces to
\[
\bar{F}=-\frac{ie^2}{8\pi^2}\int d{\bf p}\frac{m\;\hat{\bf k}
	\cdot{\bf e}\times{\bf e'}}{k_0[(\hat{\bf k}
		\cdot{\bf p})^2-p_0^2]}\; \phi(p), 
\]
and, after angular integrations,
\begin{equation}
\bar{F}=-\frac{ie^2 \hat{\bf k}\cdot{\bf e}\times{\bf e'}}{4\pi 
	k_0}I_1,\label{simp}
\end{equation}
where
\begin{equation}
I_1=\int_0^\infty dp \frac{mp}{p_0}\log\left|\frac{p_0-p}{p_0+p}
	\right| \phi(p).
\end{equation}

	With the substitution of (\ref{simp}) in (\ref{gamma}), we have
\begin{equation}
\Gamma(\spect{1}{S}{0}\rightarrow\gamma\gamma)=
	\frac{\alpha^2}{(2\pi)^2}\int d\Omega_{{\bf k}}\frac{1}{2}
	\left|\hat{\bf k}\cdot{\bf e}\times{\bf e'}\right|^2|I_1|^2 ,
\end{equation}
and, upon summation over the final polarization states,
\begin{equation}
\sum_{pol} \left|\hat{\bf k}\cdot{\bf e}\times{\bf e'}\right|^2=2,
\end{equation}
while, in view of the indistinguishability of the two photons,
\begin{equation}
\int d\Omega_{\bf k} = 2\pi.
\end{equation}
Thus, the decay rate is given by
\begin{equation}
\Gamma(\spect{1}{S}{0}\rightarrow\gamma\gamma)=
	\frac{\alpha^2}{2\pi}|I_1|^2 .
\end{equation}
This agrees with the result in Ref.~\cite{ackleh}, where the authors
have obtained the decay rate for the $\spect{1}{S}{0}$ state but not
for the $\spect{3}{P}{0}$ and $\spect{3}{P}{2}$ states.

\subsection{$\spect{3}{P}{0}$ decay}

	For the $\spect{3}{P}{0}$ state (\ref{3p0}), the decay 
amplitude (\ref{amp2}) becomes
\begin{equation}
\bar{F}=\frac{i}{(2\pi)^{3/2}}\int d{\bf p} \frac{1}{(8\pi)^{1/2}} 
	Tr[\bbox{\sigma}\cdot{\bf \hat{p}} O] \phi(p),
\end{equation}
where only terms linear in $\bbox{\sigma}$ in $O$ contribute to the
trace.  After trace evaluation and simplification, we find
\begin{equation}
\bar{F}=-\frac{ie^2}{8\pi^2} \int d{\bf p} \frac{m}{pp_0k_0^2}
	\frac{k_0({\bf e}\cdot{\bf e'})(\hat{\bf k}
	\cdot{\bf p})^2+2p_0({\bf p}\cdot{\bf e})
	({\bf p}\cdot{\bf e'})}{(\hat{\bf k}\cdot{\bf p})^2-p_0^2}, 
\end{equation}
and, upon angular integrations with the help of (\ref{angint1}),
\begin{equation}
\bar{F}=-\frac{ie^2}{4\pi k_0^2}({\bf e}\cdot{\bf e'}) I_2,\label{f3p0}
\end{equation}
where
\begin{equation}
I_2=\frac{1}{2\pi}\int_0^\infty dp \frac{mp}{p_0}[k_0A_1
	+(k_0+2p_0)A_2]\phi(p),
\end{equation}
and $A_1$ and $A_2$ are given by (\ref{angcoef1}).

	Substituting (\ref{f3p0}) in (\ref{gamma}), and 
summing over the final polarization states, we arrive at the decay rate
\begin{equation}
\Gamma(\spect{3}{P}{0}\rightarrow\gamma\gamma)=
	\frac{\alpha^2}{2\pi k_0^2}|I_2|^2.
\end{equation}

\subsection{$\spect{3}{P}{2}$ decay}

	For the $\spect{3}{P}{2}$ state (\ref{3p2}), the decay 
amplitude (\ref{amp2}) is
\begin{equation}
\bar{F}=\frac{i}{(2\pi)^{3/2}}\int d{\bf p} \left(
	\frac{3}{8\pi}\right)^{1/2}
	Tr[(\sigma_i\xi_{ij}^M\hat{p}_j) O] \phi(p),
\end{equation}
where again only terms linear in $\bbox{\sigma}$ in $O$ 
contribute to the trace.

After trace evaluation, angular integrations with the help of
(\ref{angint1}), (\ref{angint2}), and (\ref{angint3}), and applications
of the relations
\begin{equation}
\delta_{ij}\xi^M_{ij}=0,\qquad (e_ie'_j-e'_ie_j)\xi^M_{ij}=0,
\end{equation}
it is found that
\begin{equation}
\bar{F}=\frac{ie^2\sqrt{3}}{8\pi^2 k_0^2}\xi_{ij}^M \left[
	(e_je'_j+e'_ie_j)I_3+({\bf e}\cdot{\bf e'})
	\hat{k}_i\hat{k}_jI_4\right],\label{f3p2}
\end{equation}
where
\begin{eqnarray}
I_3&=&\int_0^\infty dp \left[-pA_2+\frac{2p}{p_0(m+p_0)}
	B_3\right]\phi(p),\\
I_4&=&\int_0^\infty dp \left[-\frac{pk_0}{p_0}(A_1+A_2)
	+\frac{k_0p}{p_0^2(m+p_0)}(B_1+5B_2+2B_3)
		+\frac{2p}{p_0(m+p_0)}B_2\right]\phi(p),
\end{eqnarray}
and $A_i$ and $B_i$ are given by (\ref{angcoef1}) and (\ref{angcoef3}).

	Furthermore, upon averaging over the initial states with
different values of $M$ by using (\ref{complete}), and summing over the
final polarization states, we obtain
\begin{equation}
\frac{1}{5}\sum_M\sum_{{\bf e},{\bf e'}} \bar{F}^\ast\bar{F} = 
	\frac{e^4}{80\pi^4k_0^4}(6|I_3|^2+|I_4-I_3|^2),
\end{equation}
which, when substituted in (\ref{gamma}), gives the decay rate
\begin{equation}
\Gamma(\spect{3}{P}{2}\rightarrow\gamma\gamma)=\frac{\alpha^2}{
	20\pi^3 k_0^2}(6|I_3|^2+|I_4-I_3|^2).
\end{equation}

\section{Quarkonium Photonic and Gluonic Widths}

	The quarkonium photonic and gluonic widths, which 
are obtainable by making the replacements (\ref{photonic}) and
(\ref{gluonic}) in the results of Sec.~III, are given by
\begin{eqnarray}
\Gamma(\spect{1}{S}{0}\rightarrow\gamma\gamma)&=&\frac{3 
	\alpha^2 e_Q^4}{2\pi}|I_1|^2 , \nonumber\\
\Gamma(\spect{1}{S}{0}\rightarrow gg)&=& \frac{\alpha_s^2}{3\pi}
	|I_1|^2 , \nonumber\\
\Gamma(\spect{3}{P}{0}\rightarrow\gamma\gamma)&=&\frac{3\alpha^2 
	e_Q^4}{2\pi k_0^2}|I_2|^2, \nonumber\\
\Gamma(\spect{3}{P}{0}\rightarrow gg)&=& \frac{\alpha_s^2}{3\pi 
	k_0^2}|I_2|^2  ,\label{decays}\\
\Gamma(\spect{3}{P}{2}\rightarrow\gamma\gamma)&=&\frac{3\alpha^2 e_Q^4}{
	20\pi^3 k_0^2}(6|I_3|^2+|I_4-I_3|^2), \nonumber\\
\Gamma(\spect{3}{P}{2}\rightarrow gg)&=&\frac{\alpha_s^2}{
	30\pi^3 k_0^2}(6|I_3|^2+|I_4-I_3|^2)  ,\nonumber
\end{eqnarray}
where 
\begin{equation}
k_0=\frac{1}{2}M_{Q\bar{Q}}.
\end{equation}

	We have computed these widths by using the wave 
functions and parameters obtained from our quantum-chromodynamic
potential model for heavy quarkonia \cite{quark}.  An essential feature
of our model is the inclusion of the one-loop radiative corrections in
the quantum-chromodynamic potential, which is known to be responsible
for the remarkable agreement between the theoretical and experimental
results for the spin splittings in the $c\bar{c}$ and $b\bar{b}$
spectra.  Another advantage of our model is that it is based on a
nonsingular form of the quarkonium potential, and thus avoids the use
of an illegitimate perturbative treatment.

	In addition to the wave functions,
the parameters used for the computation of the widths are
\begin{eqnarray}
\alpha_s(c\bar{c})&=&0.316,\quad m_c=2.088{\rm\ GeV},\nonumber\\
M(\eta_c)&=&2.979{\rm\ GeV}, \quad M(\chi_{c0})=3.415{\rm\ GeV}, \quad
	M(\chi_{c2})=3.556{\rm\ GeV}, \label{ccparams}
\end{eqnarray}
and
\begin{eqnarray}
\alpha_s(b\bar{b})&=&0.283,\quad m_b=5.496{\rm\ GeV},\nonumber\\
M(\eta_b)&=&9.417{\rm\ GeV}, \quad M(\chi_{b0})=9.860{\rm\ GeV}, \quad
	M(\chi_{b2})=9.913{\rm\ GeV}, \label{bbparams}
\end{eqnarray}
where we have included our theoretical value for the mass of the
unobserved energy level $\eta_b$.  Our results for $c\bar{c}$ and
$b\bar{b}$, together with the available experimental
data\cite{pdg,arm}, are given in Table~I.

\section{Alternative Treatment of Bound-State Decays}

Finally, we shall explore an alternative treatment of the bound-state
decays by making an assumption regarding the role of the potential
energy which differs from that in Sec.~II.

	Let us again consider the Fourier decomposition
\begin{equation}
\Psi({\bf r})=\frac{1}{(2\pi)^{3/2}}\int d{\bf p} 
	\Psi({\bf p})e^{i{\bf p}\cdot{\bf x}}
\end{equation}
of a bound-state $\Psi({\bf r})$, and look upon it as a superposition
of plane waves of different momenta but the same energy.  Such a
viewpoint is appropriate for a bound state because a wave packet
consisting of waves of the same energy does not spread in time.  It
also allows us to treat the decay of the ${\bf p}$ component of a bound
state of $c\bar{c}$ or $b\bar{b}$ as the annihilation of a pair of free
quark and antiquark of momenta ${\bf p}$ and $-{\bf p}$, whose energy
and effective mass are
\begin{equation}
p_0=\frac{1}{2}M_{Q\bar{Q}},\label{altenergy2}
\end{equation}
and
\begin{equation}
m=(p_0^2-{\bf p}^2)^{1/2}=\left(\frac{1}{4}M^2_{Q\bar{Q}}
	-{\bf p}^2\right)^{1/2}.
\label{mass}
\end{equation}
This approach implies that the quark and antiquark in a quarkonium can
be looked upon as free particles of constant energy and variable
momentum and mass.

	The above treatment also leads to the quarkonium decay 
rates given by (\ref{decays}) in Sec.~IV, but with an important
difference.  In Sec.~IV, $m$ is a constant, while $p_0$ is a variable,
given by (\ref{energy}).  On the other hand, in the alternative
treatment $p_0$ is a constant, while $m$ is a variable, and they are
given by (\ref{altenergy2}) and (\ref{mass}).  The computed photonic
and gluonic widths resulting from the alternative treatment are also
given in Table~I.

\section{Discussion}

	We have obtained the two-photon and the two-gluon relativistic decay 
rates of $c\bar{c}$ and $b\bar{b}$ by using two different approaches,
which are based on apparently reasonable but very different
assumptions.  It is interesting to find that both approaches give quite
similar results. As shown in Table~I, our results for the $\spect{3}{P}{0}$ and
$\spect{3}{P}{2}$ states are in agreement with the Particle Data Group
\cite{pdg}, while our results for the two-gluon decay of the
$\spect{1}{S}{0}$ state disagree
with the Particle Data Group but agree with the more recent findings of
the E760 collaboration \cite{arm}.

	In Table~I, we have also included the nonrelativistic results
obtained in Ref.~\cite{olsson} with the use of the Cornell potential.
The nonrelativistic decay rates are much smaller than the experimental
values, and this disagreement has not been
resolved by the authors in Refs.~\cite{suaya,berg,li,ackleh}, who found
that the relativistic corrections amount to a reduction in the
nonrelativistic decay rates.

	Our treatment differs from those of the earlier authors in several
respects:
\begin{enumerate}
\item	We have used totally relativistic decay amplitudes instead
of making nonrelativistic approximations or retaining only the
leading relativistic corrections.
\item	We have used a sophisticated quarkonium potential instead of
simpler potentials such as the harmonic oscillator or the Cornell potential.
\item	We have used a nonperturbative treatment for the spin-dependent
interaction terms in the quarkonium potential instead of obtaining
their contributions to the energy levels through first-order perturbation.
\end{enumerate}

	It is interesting that our values of the heavy quark masses, given
by (\ref{ccparams}) and (\ref{bbparams}), are somewhat higher than those
generally found in the literature.  This is a consequence of the fact that
nonperturbative treatment of the spin-dependent interaction terms in
the quarkonium potential yields
constituent quark masses which are higher than those resulting from 
the commonly used perturbative treatment\cite{note}. 
The nonperturbative treatment also
has a pronounced effect on the quarkonium wave functions.

	The remarkable
agreement between our theoretical results and the experimental data 
represents a distinct success of our relativistic
treatment of the bound-state decays as well as of our
quantum-chromodynamic quarkonium model.  The procedures and
mathematical techniques used in this paper can also be applied to other
decay processes.

\acknowledgments
	This work was supported in part by the U.S. Department of Energy
under Grant No.~DE-FG02-85ER40209 and the National Science Foundation
under Grant No.~PHY-93-07980.

\appendix
\section{Fermion-Antifermion Wave Functions}

	The fermion-antifermion wave functions in the matrix 
representation are given by
\begin{eqnarray}
\Psi(\spect{1}{S}{0})&=&\sqrt{\frac{1}{8\pi}}\sigma_2 \phi(r),\\
\Psi(\spect{3}{S}{1})&=&\sqrt{\frac{1}{8\pi}}\sigma_i\xi^M_i
	\sigma_2 \phi(r),\\
\Psi(\spect{3}{P}{0})&=&\sqrt{\frac{1}{8\pi}}\sigma_i\hat{x}_i
	\sigma_2 \phi(r),\\
\Psi(\spect{3}{P}{1})&=&\sqrt{\frac{3}{16\pi}}\epsilon_{ijk}\sigma_i
	\xi^M_j\hat{x}_k\sigma_2 \phi(r),\\
\Psi(\spect{3}{P}{2})&=&\sqrt{\frac{3}{8\pi}}\sigma_i
	\xi^M_{ij}\hat{x}_j\sigma_2 \phi(r),\\
\Psi(\spect{1}{P}{1})&=&\sqrt{\frac{3}{8\pi}}\xi^M_i\hat{x}_i
	\sigma_2 \phi(r),
\end{eqnarray}
where $\xi^M_i$ is a unit vector, and $\xi^M_{ij}$ is a symmetric and
traceless unit tensor, such that
\begin{equation}
\sum_{M=-1}^1 {\xi^M_i}^\star \xi^M_j = \delta_{ij},
\end{equation}
and
\begin{equation}
\sum_{M=-2}^2 {\xi^M_{ij}}^\ast \xi^M_{mn} = \frac{1}{2}(\delta_{im}
	\delta_{jn}+\delta_{in}\delta_{jm})-\frac{1}{3}\delta_{ij}
	\delta_{mn} .\label{complete}
\end{equation}
It is also to be noted that since we treat the spin-dependent
interaction terms nonperturbatively, the radial wave function
$\phi(r)$ has a different form for each of the above states.

	It is straightforward to show that these wave functions 
have the desired quantum numbers by applying the operators $L_i^2$,
$S_i^2$, and $J_i^2=L_i^2+S_i^2+2L_iS_i$ to them, and keeping in mind
that in the matrix representation of the wavefunctions \begin{equation}
S_i\Psi=\frac{1}{2}(\sigma_i\Psi + \Psi\sigma_i^T).  \end{equation}

	In the momentum space, the corresponding wave functions are
\begin{eqnarray}
\Psi(\spect{1}{S}{0})&=&\sqrt{\frac{1}{8\pi}}\sigma_2 \phi(p),
	\label{1s0}\\
\Psi(\spect{3}{S}{1})&=&\sqrt{\frac{1}{8\pi}}\sigma_i\xi^M_i
	\sigma_2 \phi(p),\\
\Psi(\spect{3}{P}{0})&=&\sqrt{\frac{1}{8\pi}}\sigma_i\hat{p}_i
	\sigma_2 \phi(p),
	\label{3p0}\\
\Psi(\spect{3}{P}{1})&=&\sqrt{\frac{3}{16\pi}}\epsilon_{ijk}\sigma_i
	\xi^M_j\hat{p}_k\sigma_2 \phi(p),\\
\Psi(\spect{3}{P}{2})&=&\sqrt{\frac{3}{8\pi}}\sigma_i
	\xi^M_{ij}\hat{p}_j\sigma_2 \phi(p),\label{3p2}\\
\Psi(\spect{1}{P}{1})&=&\sqrt{\frac{3}{8\pi}}\xi^M_i\hat{p}_i
	\sigma_2 \phi(p),
\end{eqnarray}
where we have abbreviated $|{\bf p}|$ as $p$.

\section{Angular Integrations of Decay Amplitudes}

	Angular integrations of complicated integrals appearing 
in the decay amplitudes can be performed by setting
\begin{eqnarray}
\int d\Omega_{\bf p}\; \frac{p_ip_j}{({\bf \hat{k}}\cdot{\bf p})^2
	-p_0^2}&=&A_1 \hat{k}_i\hat{k}_j +A_2 \delta_{ij},
	\label{angint1}\\
\int d\Omega_{\bf p}\; \frac{p_ip_jp_k}{({\bf \hat{k}}\cdot{\bf p})^2
	-p_0^2}&=&0,\label{angint2}\\
\int d\Omega_{\bf p}\; \frac{p_ip_jp_kp_l}{({\bf \hat{k}}
		\cdot{\bf p})^2-p_0^2}&=&
	B_1 \hat{k}_i\hat{k}_j\hat{k}_k\hat{k}_l 
	+B_2 (\delta_{ij}\hat{k}_k\hat{k}_l +\delta_{jk}\hat{k}_i
		\hat{k}_l+\delta_{ik}\hat{k}_j\hat{k}_l 
		+\delta_{il}\hat{k}_j\hat{k}_k\nonumber\\
& &		+\delta_{jl}\hat{k}_i\hat{k}_k 
		+\delta_{kl}\hat{k}_i\hat{k}_j)
		+B_3(\delta_{ij}\delta_{kl}+\delta_{ik}\delta_{jl}
		+\delta_{il}\delta_{jk}) .
	\label{angint3}
\end{eqnarray}

	Then, $A_i$ and $B_i$ are found to be
\begin{eqnarray}
A_1&=& \frac{\pi}{p_0 p}\left[6p_0p+(3p_0^2-p^2)
	\log\left|\frac{p_0-p}{p_0+p}\right|\;\right],\nonumber\\
A_2&=& -\frac{\pi}{p_0 p}\left[2p_0p+(p_0^2-p^2)
	\log\left|\frac{p_0-p}{p_0+p}\right|\;\right];
		\label{angcoef1}\\
B_1&=& \frac{\pi}{12p_0 p}\left[210p_0^3p-110p_0p^3
		+(105p_0^4-90p_0^2p^2+9p^4)
	\log\left|\frac{p_0-p}{p_0+p}\right|\;\right],\nonumber\\
B_2&=& -\frac{\pi}{12p_0 p}\left[30p_0^3p-26p_0p^3
		+(15p_0^4-18p_0^2p^2+3p^4)
	\log\left|\frac{p_0-p}{p_0+p}\right|\;\right],\nonumber\\
B_3&=& \frac{\pi}{12p_0 p}\left[6p_0^3p-10p_0p^3+3(p_0^2-p^2)^2
	\log\left|\frac{p_0-p}{p_0+p}\right|\;\right].\label{angcoef3}
\end{eqnarray}

\newlength{\jims}\setlength{\jims}{5.0in}
\begin{table}
\caption{Photonic and gluonic widths of $c\bar{c}$ and $b\bar{b}$.  The
first two sets of theoretical results correspond to the relativistic
treatments of Sec.~IV and Sec.~V.  We also give the nonrelativistic
results from Ref.~\protect\cite{olsson}.
The experimental results for $\chi_{c0}$ and $\chi_{c2}$
are from Ref.~\protect\cite{pdg}, and those for $\eta_c$ are from
Ref.~\protect\cite{arm}. }
\bigskip
\centerline{ \parbox{5.0in}{
\renewcommand{\baselinestretch}{1}
\begin{tabular}{rcccr@{}l}
Decay&Theory& Alternative& Nonrelativistic&\multicolumn{2}{c}{Expt.}\\
	&	&	Theory&	Theory&	\\
\tableline
$\eta_c\rightarrow \gamma\gamma$& 10.94 keV&	10.81 keV& &
	6.7&\mbox{\renewcommand{\baselinestretch}{0.5}%
	$\scriptsize\begin{array}{l}+2.4\\-1.7\end{array}$}$\pm 2.3$ keV\\
$\rightarrow gg$&		23.03 MeV&	22.76 MeV& 9.01 MeV&
	23.9&\mbox{\renewcommand{\baselinestretch}{0.5}%
	$\scriptsize\begin{array}{l}+12.6\\-7.1\end{array}$} MeV\\
$\chi_{c0}\rightarrow \gamma\gamma$& 6.38 keV&	8.13 keV& &
	4.0&$\pm 2.8$ keV\\
$\rightarrow gg$&		13.44 MeV&	17.10 MeV& 1.63 MeV&
	13.5&$\pm 3.3\pm 4.2$ MeV\\
$\chi_{c2}\rightarrow \gamma\gamma$& 0.57 keV&	1.14 keV& &
	0.321&$\pm 0.078\pm 0.054$ keV\\
$\rightarrow gg$&		1.20 MeV&	2.39 MeV& 0.37 MeV&
	2.00&$\pm 0.18$ MeV\\
$\eta_b\rightarrow \gamma\gamma$&0.46 keV&	0.48 keV& &
	\multicolumn{2}{c}{ }\\
$\rightarrow gg$&		12.46 MeV&	13.02 MeV& &
	\multicolumn{2}{c}{ }\\
$\chi_{b0}\rightarrow \gamma\gamma$&0.080 keV&	0.085 keV& &
	\multicolumn{2}{c}{ }\\
$\rightarrow gg$&		2.15 MeV&	2.29 MeV& &
	\multicolumn{2}{c}{ }\\
$\chi_{b2}\rightarrow \gamma\gamma$&0.008 keV&	0.012 keV& &
	\multicolumn{2}{c}{ }\\
$\rightarrow gg$&		0.22 MeV&	0.33 MeV& &
	\multicolumn{2}{c}{ }	 
\end{tabular} }}
\end{table}

\begin{thebibliography}{99}
\bibitem{note}  We have checked, for instance, that if the hyperfine
	interaction terms in the quarkonium potential are treated by
	means of first-order perturbation, the charmed and bottom quark
	masses required to fit the experimental data for the
	hyperfine splittings in $c\bar{c}$ and $b\bar{b}$ are
	$m_c=1.58$~GeV and $m_b=5.1$~GeV, which are lower than the values
	given by Eqs.~(\ref{ccparams}) and (\ref{bbparams}) in
	Sec.~IV.
\end{thebibliography}
\end{document}